\documentclass[runningheads]{llncs}

\usepackage[numbers]{natbib} 
\usepackage[T1]{fontenc}
\usepackage{tabularx}
\usepackage{subfiles}
\usepackage{hyperref}
\usepackage{balance}
\usepackage{booktabs}
\usepackage{natbib}
\usepackage{caption}
\usepackage{graphicx}
\usepackage{subcaption} 
\usepackage{adjustbox} 

\begin{document}
%
\title{Advancing User-Voice Interaction: Exploring Emotion-Aware Voice Assistants Through a Role-Swapping Approach}
%
%
\author{
Yong Ma\inst{1}\orcidID{0000-0002-8398-4118} \and
Yuchong Zhang\inst{2}\orcidID{0000-0003-1804-6296} \and
Di Fu\inst{3}\orcidID{0000-0002-5385-2982} \and
Stephanie Zubicueta Portales\inst{4}\orcidID{0009-0004-2459-1949} \and
Danica Kragic\inst{2}\orcidID{0000-0003-2965-2953} \and
Morten Fjeld\inst{1,5}\orcidID{0000-0002-9562-5147}
}

\authorrunning{Y. Ma et al.}
%
\institute{University of Bergen, Bergen, Norway 
\and KTH Royal Institute of Technology, Stockholm, Sweden \and University of Surrey, Guildford, UK \and Chalmers University of Technology, Gothenburg, Sweden \and Norwegian University of Science and Technology,Trondheim, Norway
\\}





%
\maketitle              
\begin{abstract}
As voice assistants (VAs) become increasingly integrated into daily life, the need for emotion-aware systems that can recognize and respond appropriately to user emotions has grown. While significant progress has been made in speech emotion recognition (SER) and sentiment analysis, effectively addressing user emotions—particularly negative ones—remains a challenge. This study explores human emotional response strategies in VA interactions using a role-swapping approach, where participants regulate AI emotions rather than receiving pre-programmed responses. Through speech feature analysis and natural language processing (NLP), we examined acoustic and linguistic patterns across various emotional scenarios. Results show that participants favor neutral or positive emotional responses when engaging with negative emotional cues, highlighting a natural tendency toward emotional regulation and de-escalation. Key acoustic indicators such as root mean square (RMS), zero-crossing rate (ZCR), and jitter were identified as sensitive to emotional states, while sentiment polarity and lexical diversity (TTR) distinguished between positive and negative responses. These findings provide valuable insights for developing adaptive, context-aware VAs capable of delivering empathetic, culturally sensitive, and user-aligned responses. By understanding how humans naturally regulate emotions in AI interactions, this research contributes to the design of more intuitive and emotionally intelligent voice assistants, enhancing user trust and engagement in human-AI interactions.

\end{abstract}

\section{Introduction}
Voice assistants (VAs) have become an integral part of our daily life, offering convenience and efficiency in tasks ranging from information retrieval to smart home control. However, traditional VAs primarily focus on functionality, such as in-home systems like Alexa~\cite{mclean2019hey,hoy2018alexa}, often overlooking the emotional nuances of human interaction. As voice user interface (VUI) technologies advance, the ability to recognize, understand, and appropriately respond to user emotions is becoming increasingly important.
Emotion-aware VAs, capable of detecting and responding to users' emotional states, have the potential to create more empathetic and engaging user experiences~\cite{ma2023emotion,kossack2023emotion}.

Recent advancements in artificial intelligence (AI), natural language processing (NLP), and affective computing have enabled VAs to detect emotional cues from speech, including vocal patterns, tone, and contextual information. This capability allows VAs to dynamically adjust their responses, tailoring interactions to the user's emotional state and improving user satisfaction and trust~\cite{ma2022should,Parvathi2025voice}.
However, while advancements in speech emotion recognition and speech emotion synthesis have enabled the recognition and understanding of user emotions~\cite{wani2021comprehensive} and the generation of emotionally expressive speech~\cite{triantafyllopoulos2023overview}, effectively responding to user emotions—particularly negative ones—remains a significant challenge.
One major obstacle lies in the variability of emotional expression across individuals and cultures. Negative emotions, such as anger or sadness, can be conveyed differently depending on cultural norms, making it difficult for systems to accurately detect and respond to these emotions~\cite{mesquita1992cultural}. For instance, speech emotion recognition (SER) systems may struggle with accents, dialects, or background noise, leading to misinterpretations of emotional cues~\cite{can2023approaches}.  

Beyond detection, responding appropriately to negative emotions is equally complex. While empathetic reactions can be beneficial to users~\cite{raamkumar2022empathetic,hu2022acoustically}, as seen in Alexa's empathetic responses~\cite{carolus2021alexa}, striking the right balance between empathy and functionality is crucial. Overly intrusive or artificial responses can feel misaligned, ultimately undermining user trust and satisfaction~\cite{atta2024influence,berking2012emotion}. Moreover, emotional expression varies based on cultural, gender, and contextual differences~\cite{fischer2004gender,fischer2000relation}. As a result, purely empathetic emotional responses may not always be suitable or effective for all users. Developing adaptive, context-aware strategies for emotion-aware systems remains a critical area for future research.

\begin{figure}[ht]
\centering
  \includegraphics[width=0.8\columnwidth]{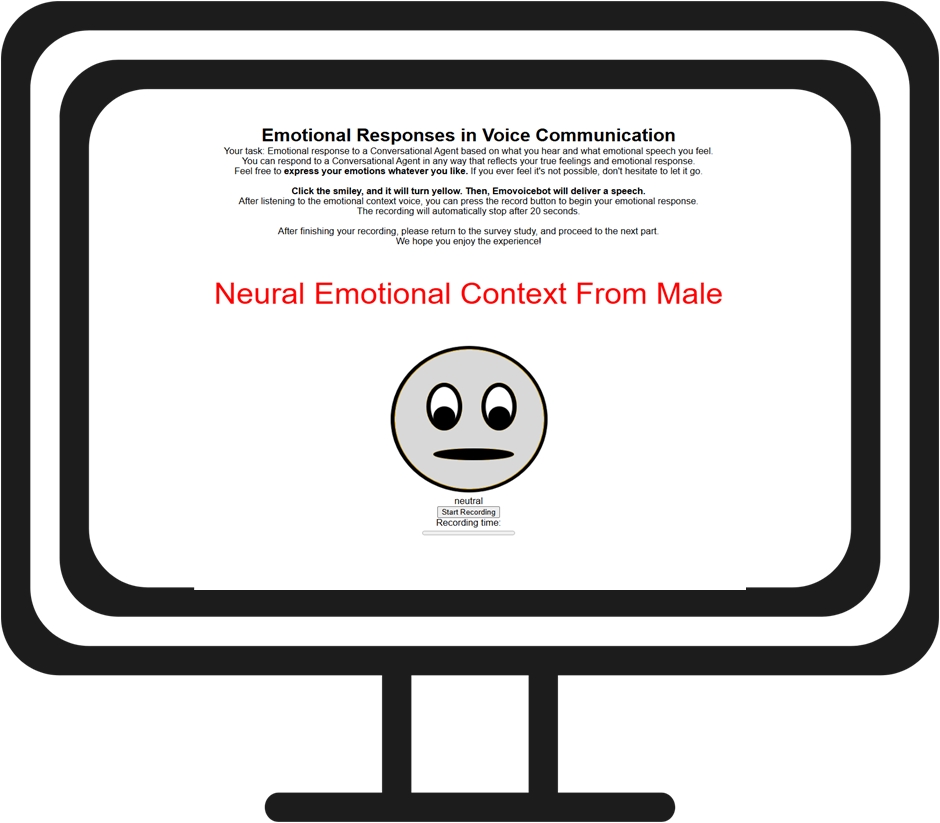}
  \caption{The web page is designed to collect voice samples from participants. When an emoji is clicked, it turns yellow and plays an emotional context (e.g., a sad or happy scenario). Participants are then prompted to say something comforting or engage in a conversation with the emoji by clicking the "Start Recording" button. This interactive design allows users to respond naturally to the emotional context, providing valuable data for emotion-aware systems.}\label{fig:webpage}
  \vspace{-1em}
\end{figure} 


To address this issue, an effective approach is to understand how people respond to different emotions in various contexts. By studying human emotional responding strategies, we can transfer these insights into voice assistants (VAs) to create more empathetic and contextually appropriate interactions. This approach aligns with the concept of role-swapping~\cite{ma2022should,huang2024relationship}, where human strategies for emotional responses are mimicked in AI systems.
In this study, we employ a role-swapping approach, where the traditional roles of VAs and individuals are reversed. Participants are tasked with applying their own strategies to respond to various emotional contexts displayed by VAs. The primary objective of this study is to influence and regulate the emoji's emotions, guiding them toward more positive or neutral states. To achieve this, participants are encouraged to engage with the VAs using diverse emotional tones and responses, allowing for dynamic and adaptive interactions.
To facilitate this, we designed a website (shown in Figure~\ref{fig:webpage}) where participants can join the study via a provided web link. On the website, participants can input their emotional responses, which are automatically recorded. After voice recording, the participants' voices are analyzed for emotion detection and speech features.
By engaging with emotionally expressive VAs, participants apply their own emotional strategies to interact in ways that aim to de-escalate or regulate emotional states. Through speech feature analysis, sentiment polarity assessments, and linguistic evaluations, our study reveals key insights into how users react to different emotional scenarios. The results show that participants favor neutral or positive emotional responses when interacting with negative emotions, indicating a natural tendency toward emotional regulation in human-AI interactions. By integrating these findings, we aim to advance the development of emotion-aware VAs, enabling more empathetic, natural, and effective interactions between humans and voice assistants.

\section{Related Work}
\subsection{Emotion Recognition in Voice Assistants}
Emotion recognition is a critical component of emotion-aware voice assistants (VAs). By analyzing speech features such as tone, pitch, and speech patterns~\cite{el2011survey,rathi2024analyzing,ma2024understanding}, speech emotion recognition (SER) enables VAs to detect emotional states from speech signals. Recent advancements in AI technologies, particularly in Large Language Models (LLMs) ~\cite{ma2024leveraging} and multimodal approaches~\cite{kumar2024multimodal} to SER, have significantly enhanced the accuracy and robustness of emotion detection ~\cite{lieskovska2021review,khalil2019speech}. These technologies leverage machine learning and deep learning models to identify emotions such as happiness, sadness, and anger, fostering more emotionally intelligent and contextually aware interactions in VAs.
Despite these advancements, challenges remain in handling variability in emotional expression due to factors such as cultural differences, accents, and background noise~\cite{mesquita1992cultural,can2023approaches}. For example, SER systems may misinterpret emotional cues in diverse linguistic or cultural contexts, reducing their effectiveness in real-world applications. Addressing these limitations requires further research into adaptive and context-aware emotion recognition models that can better generalize across different user demographics and environmental conditions.

\subsection{Empathetic Responses in Human-Computer Interaction}
Empathy in AI refers to a conversational agent's ability to recognize, interpret, and respond appropriately to a user’s emotional state~\cite{srinivasan2022role}. Empathetic responses play a crucial role in fostering engaging and emotionally resonant interactions\cite{raamkumar2022empathetic,liu2022artificial}, enhancing user engagement and trust by adapting speech tone, word choices, and emotional expressiveness\cite{barange2022impact}.
studies have shown that users respond positively to empathetic reactions from AI systems, such as Alexa’s ability to provide comforting and supportive responses\cite{carolus2021alexa,mari2024empathic}. However, designing appropriate responses to negative emotions remains a significant challenge. Although empathetic reactions can improve user satisfaction, overly intrusive or artificial responses may feel misaligned with user expectations, ultimately undermining trust\cite{atta2024influence,berking2012emotion}. This underscores the need for contextually appropriate and culturally sensitive emotional responses in emotion-aware AI systems.


\subsection{Role-Swapping in Emotion Regulation}
Role-swapping in emotion regulation is a promising approach for enhancing voice user interface design, particularly in the development of emotion-aware voice assistants (VAs). By reversing traditional roles, users take on the responsibility of regulating the emotions of AI systems, such as responding to negative emotional cues displayed by VAs. This approach allows researchers to explore alternative emotional responding and reacting strategies, shedding light on how humans manage AI-driven emotions.
Studies have shown that individuals often adopt neutral emotional responses when interacting with emotionally expressive AI systems. However, gender and cultural differences can significantly influence these strategies~\cite{ma2022should,huang2024relationship}. Furthermore, research on interpersonal emotion regulation highlights how individuals adjust their strategies to support others—for example, using cognitive reappraisal to reinterpret stressful situations or employing expressive suppression to prevent conflict escalation~\cite{zaki2013interpersonal,gross2015emotion}. These insights can be applied to AI systems, enabling them to mimic human emotional responding strategies and create more empathetic and adaptive interactions.
Despite its potential, role-swapping in emotion regulation presents challenges, particularly in addressing cultural and contextual differences in emotional expression and ensuring that such systems remain adaptive and effective across diverse user demographics~\cite{mesquita1992cultural,fischer2004gender}. By leveraging role-swapping approaches, researchers can develop emotion-aware VAs capable of providing contextually appropriate and emotionally resonant responses, ultimately enhancing user satisfaction and trust in AI-driven interactions.

\subsection{Research Gaps}
While significant progress has been made in emotion recognition and empathetic response generation, research on how humans naturally respond to emotional cues from AI systems—particularly in role-swapping scenarios—remains limited. Most existing studies focus on how AI systems respond to human emotions, rather than how humans regulate the emotions of AI systems.
Our study addresses this gap by introducing a role-swapping approach, where participants interact with emotion-aware voice assistants (VAs) to transform negative emotions into positive or neutral states. By analyzing participants’ emotional responses, we aim to develop effective strategies for designing more adaptive and empathetic VAs. This approach not only deepens our understanding of human emotional responding strategies but also provides valuable insights for enhancing emotion-aware AI systems, leading to more natural and engaging human-AI interactions.
\section{Experiment Setup}
As shown in Figure~\ref{fig:webpage}, we conducted an online user study to explore users' strategies for responding to different emotional contexts. By analyzing participants' voices and the content of their interactions with our one-dialog simulated VAs, we evaluated emotional responding differences, speech features, and language features using speech analysis and natural language processing (NLP) techniques. This approach allowed us to gain insights into how users naturally adapt their responses to emotional cues, providing valuable data for improving emotion-aware systems.

\subsection{Emotional Scenarios}
In our study, we designed five distinct emotional scenarios based on the five basic emotions — neutral, happy, sad, angry, and fear~\cite{ortony1990s,ortony2022all,ma2022emotion}. These scenarios were crafted to reflect situations that users might encounter in their daily lives, ensuring they were relatable and easily identifiable by general users. After confirming the emotional states, we conducted brainstorming sessions to determine which scenarios would best represent these emotions in a way that felt authentic and meaningful. The final five emotional scenarios are as follows:

\begin{itemize}
    \item \textbf{Neutral:} I put on my shoes before leaving the house.
    \item \textbf{Happy:} I am visiting my favorite country or city.
    \item \textbf{Sad:} I see children suffering from disease, sickness, or war.
    \item \textbf{Angry:} I get betrayed by a close friend or relative.
    \item \textbf{Fear:} I am walking in the dark in the woods when I stumble upon a dead body. The blood seems fresh, and I hear a branch breaking from behind.
\end{itemize}

Based on these emotional scenarios, we recorded voice clips for each scenario and integrated them into our designed website. Users visiting the website can listen to these emotional scenario recordings, allowing them to immerse themselves in the context and respond naturally. This interactive setup enables participants to engage with the emotional content and provide authentic responses, which are then recorded and analyzed for further study.

\subsection{Apparatus, Participants and Experimental Procedure}
\subsubsection{Apparatus and Participants}
In this user study, we primarily utilized OpenVokaturi\footnote{\url{https://developers.vokaturi.com/getting-started/overview}} to analyze the emotional states of speech. Additionally, we employed two key speech analysis packages: Librosa\footnote{\url{https://librosa.org/doc/latest/feature.html}} and OpenSMILE\footnote{\url{https://audeering.github.io/opensmile-python/}}, to examine users' speech behaviors in response to emotional scenarios. For analyzing the content of users' responses, we relied on four main NLP packages: WordCloud\footnote{\url{https://amueller.github.io/word_cloud/}}, NLTK\footnote{\url{https://www.nltk.org/}}, TextStatc, and SpaCy\footnote{\url{https://spacy.io/}}. Furthermore, OpenAI Whisper\footnote{\url{https://openai.com/index/whisper/}} was used as a critical tool for converting speech into text, enabling detailed analysis of both linguistic and emotional features. Participants' voices were recorded using the built-in microphones of their own computers and uploaded to our web server when they conducted our user study on our website. All recorded voice signals were captured at a sample rate of 48 kHz, ensuring high-quality audio for subsequent analysis.
Additionally, we recruited 60 participants through Prolific\footnote{\url{https://www.prolific.com/}}, comprising 30 males and 30 females. To account for potential gender-based differences in voice perception, we conducted two separate user studies: one using a male emotional voice and the other using a female emotional voice for the five emotional scenarios. The male emotional voice had a mean age of 32.29 years (SD = 9.71), while the female emotional voice had a mean age of 27.17 years (SD = 5.02). This approach ensured a balanced and comprehensive evaluation of emotional responses across different gender contexts.
After completing the user study, each participant was compensated with £4.50 through the Prolific website. Participants were assured of complete anonymity, as clearly stated in the website's privacy policy. This policy also detailed the types of data collected during the study, which included voice recordings and responses to demographic questions, such as age and gender. All measures were taken to ensure the confidentiality and privacy of the participants' information.

\subsubsection{Experimental Procedure}
As mentioned earlier, we recruited participants through the Prolific website, and they accessed our designed website (shown in Figure\ref{fig:webpage}) via a provided link. The landing page provided an overview of the study, along with detailed instructions. Participants could click on the smiley emoji, which would turn yellow and play an emotional scenario voice. They were then free to respond in any way they liked. When participants clicked the "\textit{Start Recording}" button, their microphone would begin recording their voice. They were given 20 seconds to speak, during which they could engage in conversation with the smiley or attempt to comfort it, especially in cases of negative emotions such as anger, fear, or sadness.
In this study, participants engaged with five different emotional scenarios, with either a male or female emotional voice randomly selected for each scenario. To ensure gender balance, 15 male and 15 female participants interacted with the male emotional voice scenarios, while the remaining 30 participants interacted with the female emotional voice scenarios. After completing all five emotional scenarios, participants finished the main user study and proceeded to fill out a relevant questionnaire.

\subsection{Speech Signals Analysis}
The speech data collected from participants were downloaded from the website server and systematically classified according to different emotional scenarios, based on the naming conventions of the audio files. To extract relevant speech features, we utilized advanced speech processing Python packages, such as Librosa and OpenSMILE~\cite{eyben2010opensmile}. Additionally, to analyze the content of the speech, we transcribed the audio data into text using OpenAI Whisper model. Leveraging natural language processing (NLP) techniques, we then extracted a variety of text features, including lexical, semantic, and syntactic features. Furthermore, we employed Openvokatuti to identify and analyze participants' emotional states during their interactions, specifically determining which emotional responses they exhibited while communicating with our VAs. This comprehensive approach allowed us to gain deeper insights into both the acoustic and linguistic aspects of the participants' interactions.


\section{Results}
Our study involved 60 participants, resulting in 242 valid audio files across five emotional scenarios: happy, neutral, angry, sad, and fear. Through a combination of speech signal analysis and linguistic evaluation, we identified key patterns in participants' emotional responses and speech features.

\subsection{Results from Speech Signals}
\subsubsection{Speech Emotion Analysis}

\begin{figure}[ht]
    \centering
    \begin{subfigure}{0.49\textwidth}
        \centering
        \includegraphics[width=\textwidth]{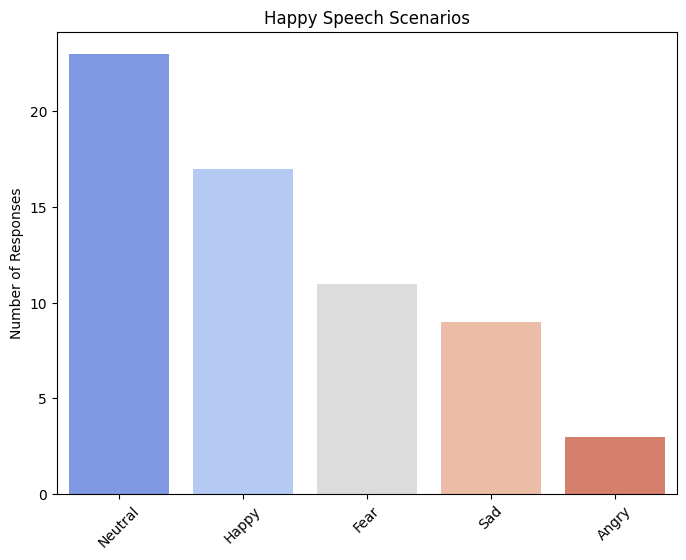}
        \caption{Happy Speech Scenarios}
        \label{fig:happy_distribution}
    \end{subfigure}
    \hfill
    \begin{subfigure}{0.49\textwidth}
        \centering
        \includegraphics[width=\textwidth]{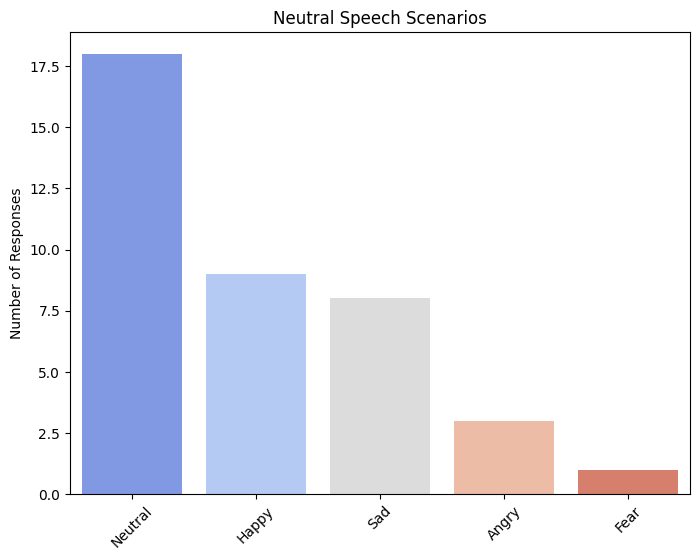}
        \caption{Neutral Speech Scenarios}
        \label{fig:neutral_distribution}
    \end{subfigure}
    
    \caption{The Emotional Responding Distribution from Different Emotional Scenarios}
    \label{fig:emotion_distribution}
\end{figure}

In this study, we aimed to uncover the strategies users employ when responding to different emotional scenarios. To achieve this, we utilized OpenVokaturi as our emotion analysis tool to extract emotional states from the collected audio data. As shown in Figure~\ref{fig:emotion_distribution}, we observed that in happy and neutral speech scenarios, participants primarily used neutral and happy emotional responses to communicate. However, in the case of negative emotional scenarios (e.g., anger, sadness, and fear), the majority of participants preferred to use neutral emotional responses when interacting with others. This suggests that users tend to adopt a balanced and non-confrontational approach when dealing with negative emotions, highlighting the importance of neutrality in emotional communication.

\subsubsection{Speech Signal Analysis}

\begin{figure}[ht]
    \centering
    \begin{subfigure}{0.32\textwidth}
        \centering
        \includegraphics[width=\textwidth]{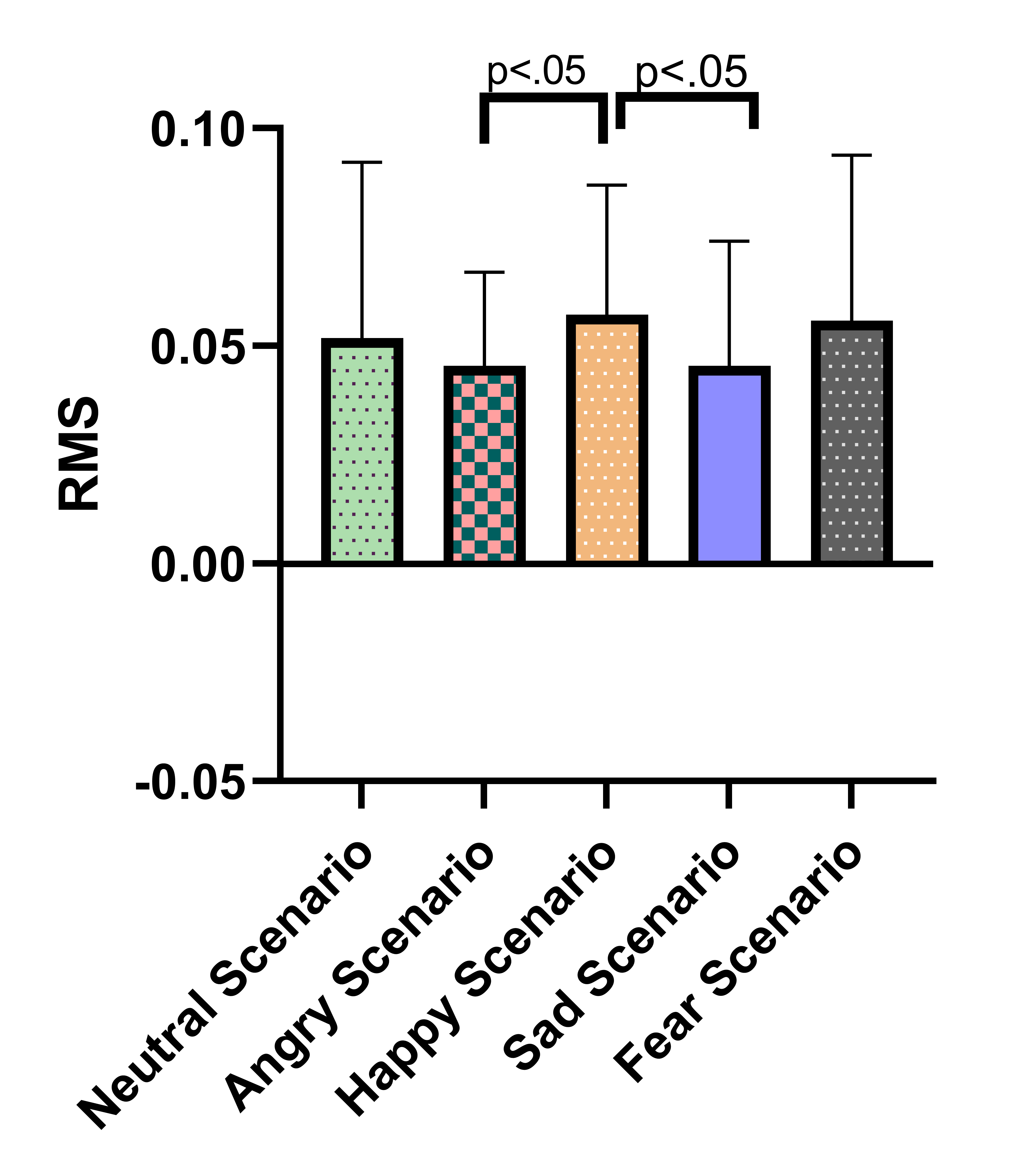}
        \caption{ Root Mean Square}
        \label{fig:rms_distribution}
    \end{subfigure}
    \hfill
    \begin{subfigure}{0.32\textwidth}
        \centering
        \includegraphics[width=\textwidth]{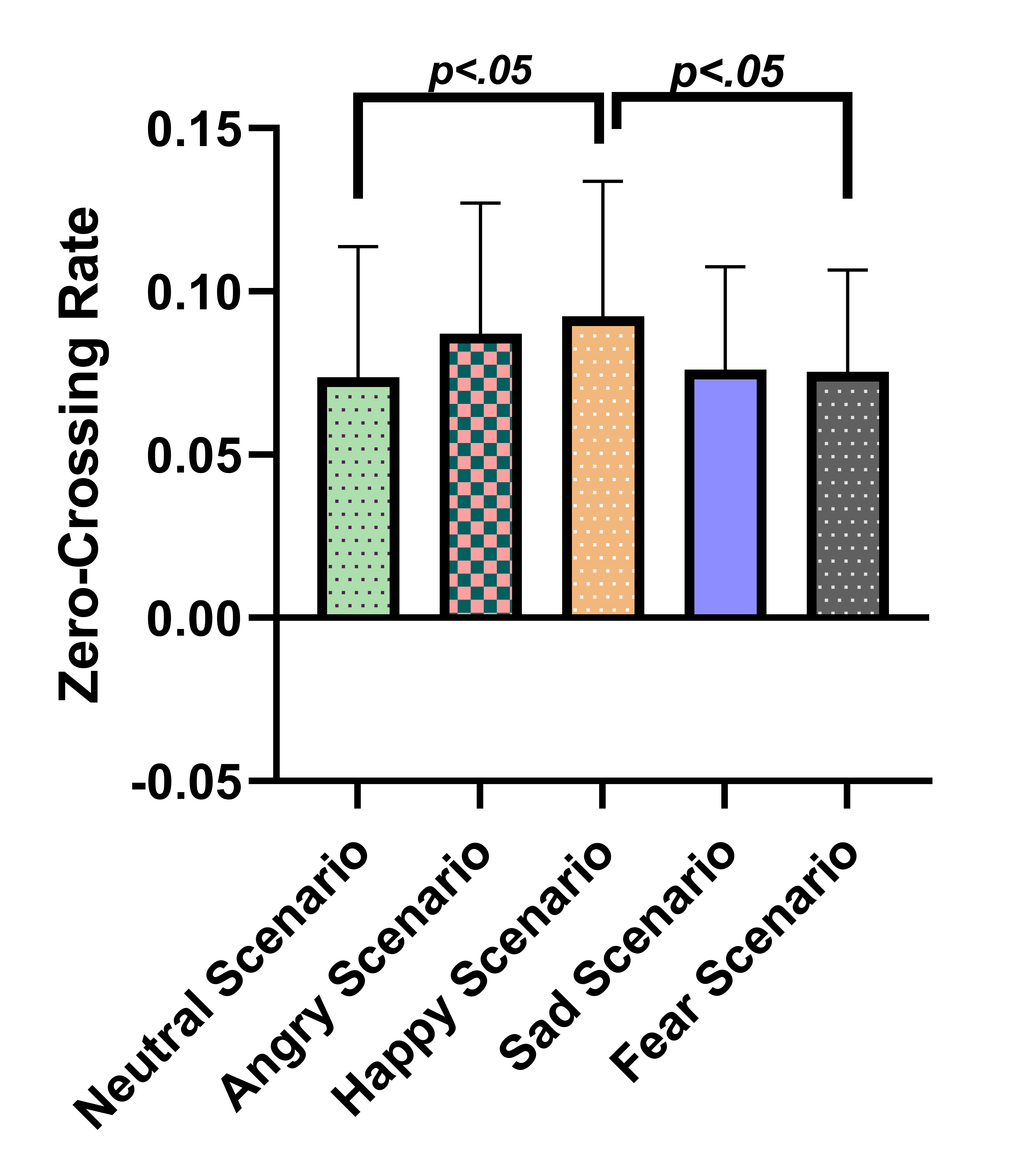}
        \caption{ Zero-Cross Rate}
        \label{fig:zcr_distribution}
    \end{subfigure}
    \hfill
    \begin{subfigure}{0.32\textwidth}
        \centering
        \includegraphics[width=\textwidth]{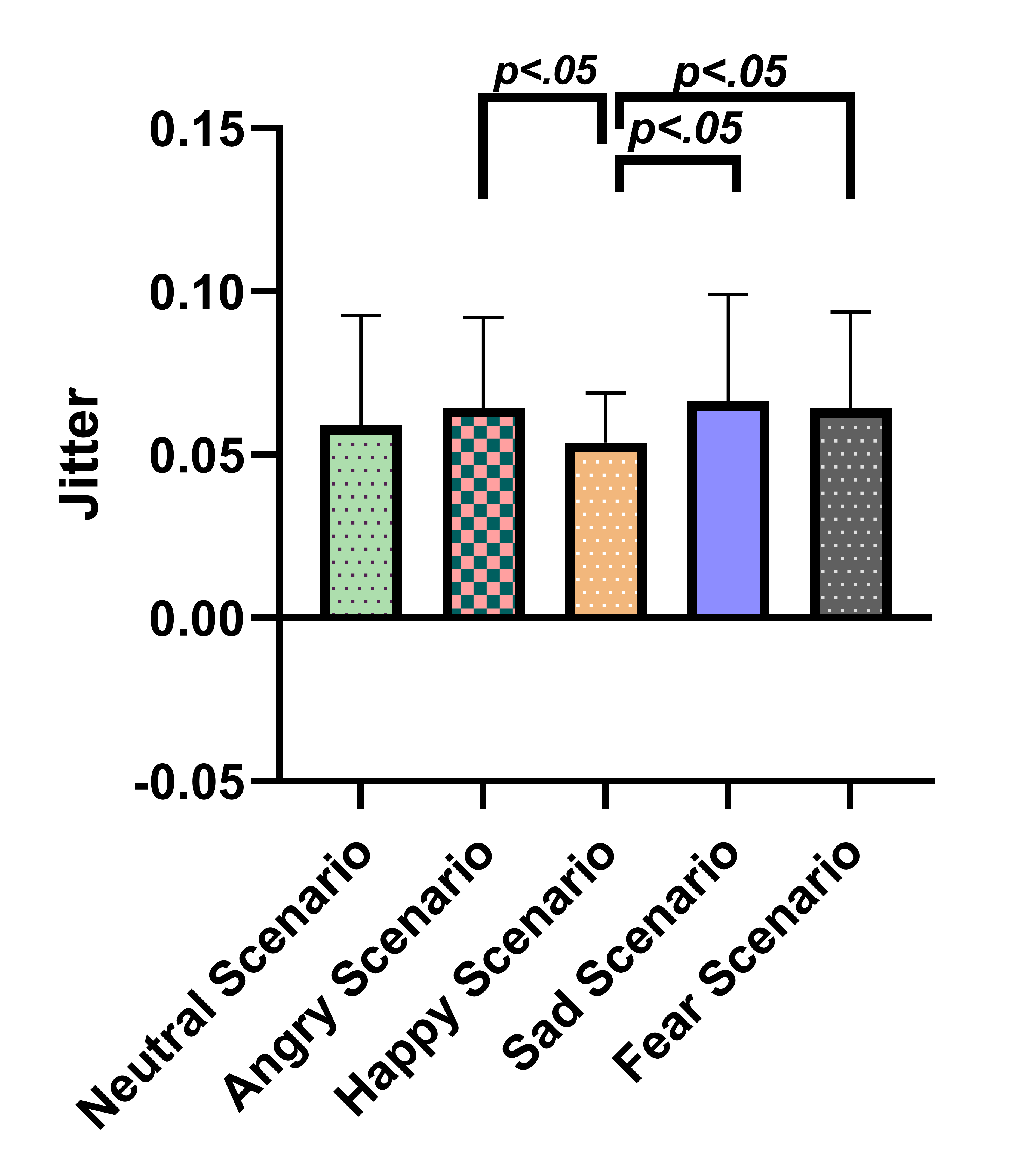}
        \caption{Jitter}
        \label{fig:jitter_distribution}
    \end{subfigure}
    
    \caption{Comparison of Different Features Using T-Test from Five Basic Emotional Scenarios: (a) Root Mean Square, (b) Zero-Cross Rate, and (c) Jitter.}
    \label{fig:feature_comparison}
\end{figure}

Speech analysis typically involves examining time-domain speech features and frequency-domain speech features (Madan et al., 2014). In this study, we primarily used Librosa and OpenSMILE to extract these features. We then employed T-tests to compare the differences between various emotional scenarios. As shown in Figure~\ref{fig:feature_comparison}, we found that only three features—root mean square (RMS), zero-crossing rate (ZCR), and speech jitter—exhibited significant differences across some emotional scenarios.
For instance, in RMS, we observed differences between the happy speech scenario and the angry speech scenario, as well as between the happy speech scenario and the sad speech scenario. In ZCR, significant differences were found between the happy speech scenario and the neutral speech scenario, as well as between the happy speech scenario and the fear speech scenario. For speech jitter, differences were identified between the happy speech scenario and the angry speech scenario, the happy speech scenario and the fear speech scenario, and the happy speech scenario and the sad speech scenario.
These findings suggest that RMS, ZCR, and jitter are sensitive to variations in emotional expression, particularly in distinguishing between positive and negative emotional states. This highlights their potential as key indicators for emotion-aware systems to detect and respond to different emotional contexts effectively.

\subsection{Results from Speech Content}
\subsubsection{Result from Word Cloud}

\begin{figure}[ht]
    \centering
    \begin{subfigure}{0.32\textwidth}
        \centering
        \includegraphics[width=\textwidth]{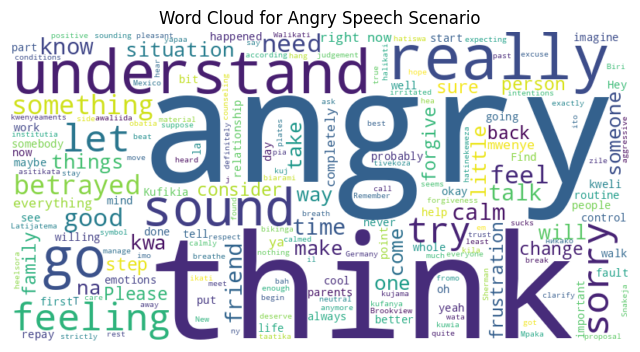}
        \caption{ Angry}
        \label{fig:wordcloud_angry}
    \end{subfigure}
    \hfill
    \begin{subfigure}{0.32\textwidth}
        \centering
        \includegraphics[width=\textwidth]{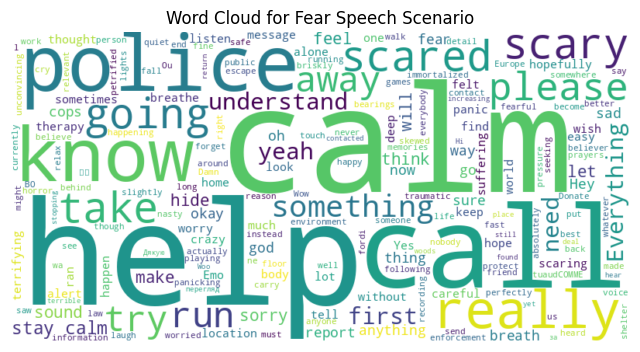}
        \caption{ Fear}
        \label{fig:wordcloud_fear}
    \end{subfigure}
    \hfill
    \begin{subfigure}{0.32\textwidth}
        \centering
        \includegraphics[width=\textwidth]{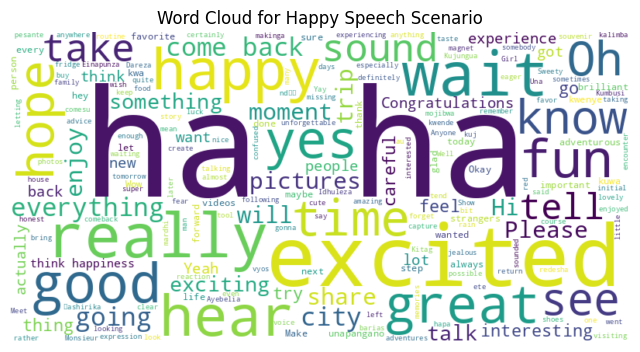}
        \caption{ Happy}
        \label{fig:wordcloud_happy}
    \end{subfigure}
    
    \vspace{0.5cm} 
    \begin{subfigure}{0.48\textwidth}
        \centering
        \includegraphics[width=\textwidth]{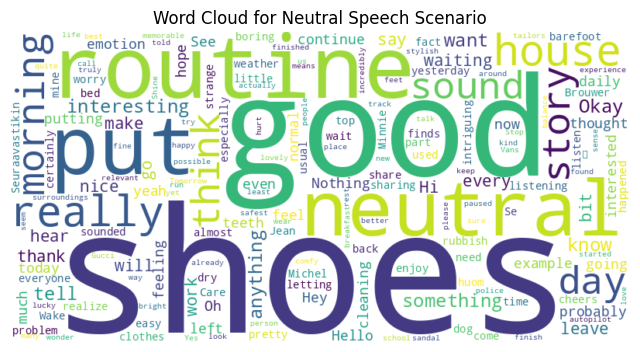}
        \caption{ Neutral}
        \label{fig:wordcloud_neutral}
    \end{subfigure}
    \hfill
    \begin{subfigure}{0.48\textwidth}
        \centering
        \includegraphics[width=\textwidth]{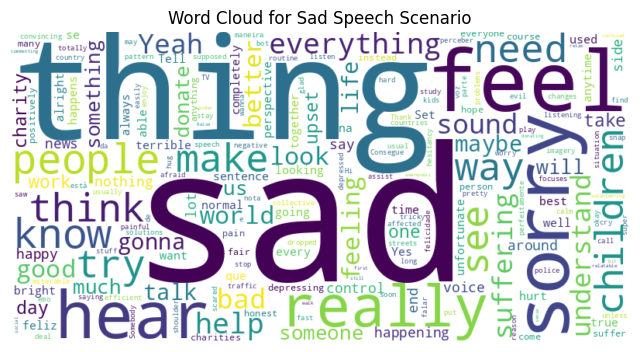}
        \caption{ Sad}
        \label{fig:wordcloud_sad}
    \end{subfigure}

    \caption{Word Cloud Representations for Different Emotional Scenarios: (a) Angry, (b) Fear, (c) Happy, (d) Neutral, and (e) Sad.}
    \label{fig:wordclouds}
\end{figure}

In this study, we utilized the wordcloud package to analyze linguistic patterns across various emotional states. The resulting word clouds, as depicted in Figure~\ref{fig:wordclouds}, provide a visual representation of how individuals linguistically express emotions in different scenarios. Each emotional state reveals unique linguistic tendencies, reflecting the underlying feelings and cognitive processes associated with anger, fear, happiness, neutrality, and sadness.
In the angry speech scenario, the most frequently used words include "angry," "think," "betrayed," and "frustration." These terms highlight a strong focus on personal reflection, conflict, and perceived injustice. The presence of words like "understand" and "really" suggests that individuals in angry states often attempt to justify their emotions or seek validation, particularly in confrontational or tense situations. This linguistic pattern underscores the cognitive effort to rationalize or communicate feelings of anger.
For the fear speech scenario, the word cloud features terms such as "help," "calm," "police," "scared," and "run." These words reflect a mix of distress and an urgent need for safety or assistance. The emphasis on phrases like "stay calm" and "know" indicates that individuals experiencing fear also strive to manage their emotions and seek solutions during stressful or threatening situations. This linguistic behavior highlights the dual focus on expressing fear while attempting to regain control.
In the happy speech scenario, the dominant words include "excited," "ha ha," "fun," "happy," and "wait." These terms reflect expressions of joy, enthusiasm, and anticipation. The frequent use of words like "see," "hear," and "trip" suggests that happiness is often associated with experiences, events, and social interactions. This linguistic pattern emphasizes the outward and experiential nature of happiness, as individuals share their positive emotions and engage with the world around them.
The neutral speech scenario is characterized by words such as "routine," "shoes," "morning," and "house." These terms indicate discussions about daily life, habits, and general observations. The frequent appearance of "story" and "interesting" suggests that even in neutral emotional states, people engage in conversations about experiences and narratives rather than expressing strong emotional reactions. This linguistic pattern reflects the mundane yet meaningful nature of everyday communication.
Finally, in the sad speech scenario, key words such as "sad," "sorry," "feel," "trying," and "children" highlight expressions of emotional distress, regret, and concern for others. Words like "understand," "hear," and "people" indicate a search for empathy, connection, or support in difficult situations. This linguistic pattern underscores the introspective and relational aspects of sadness, as individuals seek to articulate their feelings and connect with others during challenging times.
These word clouds illustrate the distinct linguistic patterns associated with different emotional scenarios, offering valuable insights into how individuals communicate their emotions. Understanding these patterns can play a crucial role in enhancing the development of emotion-aware VAs, enabling them to generate more contextually appropriate and empathetic responses tailored to users' emotional states.

\subsubsection{Result from Correlation Between NLP Features}
\begin{figure}[ht]
    \centering
    \includegraphics[width=0.7\textwidth]{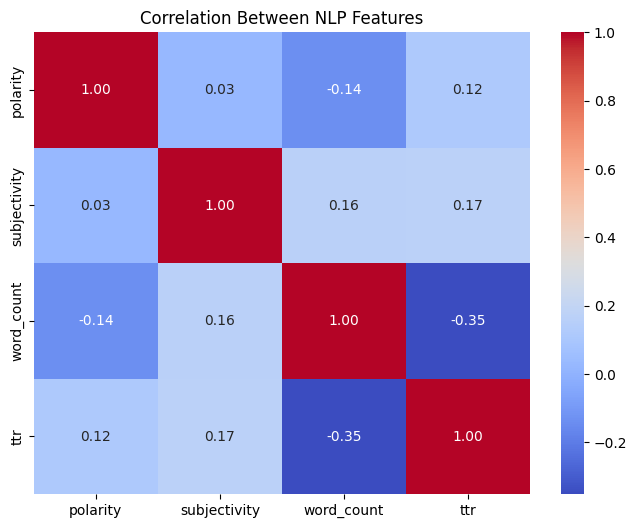} 
    \caption{Correlation Heatmap Between NLP Features. The heatmap visualizes the relationships among key linguistic features, including polarity, subjectivity, word count, and type-token ratio (TTR). Warmer colors indicate stronger positive correlations, while cooler colors represent negative correlations.}
    \label{fig:nlp_heatmap}
\end{figure}

As it show in Figure~\ref{fig:nlp_heatmap}, it reveals relationships between key linguistic metrics, including polarity, word count, and Type-Token Ratio (TTR). Polarity, which quantifies sentiment on a scale from negative to positive, exhibits a weak positive correlation (0.12) with word count, suggesting that longer texts may slightly amplify sentiment intensity.
TTR, a measure of lexical diversity, demonstrates a strong correlation with itself (1.00) and a moderate positive correlation with word count (0.16), indicating that longer texts tend to exhibit greater vocabulary variety. Notably, a significant negative correlation (-0.35) between polarity and another linguistic feature suggests an inverse relationship, potentially reflecting how emotions or linguistic structures influence sentiment expression.
These findings highlight the intricate interplay among linguistic features, offering valuable insights for the future design of emotion-aware VAs that can better interpret and respond to user sentiment.

\subsubsection{Result from Polarity and TTR Across Different Emotion Responses}
\begin{figure}[ht]
    \centering
    \includegraphics[width=0.8\textwidth]{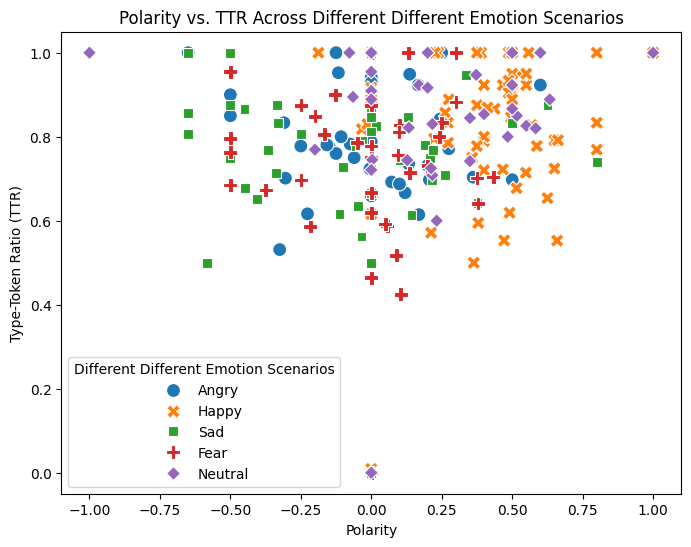}
    \caption{Polarity vs. Type-Token Ratio (TTR) Across Different Emotion Scenarios. Each marker represents a distinct emotional response: Angry, Happy, Sad, Fear, and Neutral. The distribution highlights differences in lexical diversity and sentiment across emotions.}
    \label{fig:polarity_ttr}
\end{figure}

From Figure~\ref{fig:polarity_ttr}, the relationship between polarity (sentiment positivity/negativity) and type-token ratio (TTR) (lexical diversity) across different emotional responses is illustrated. Each emotion category — Angry, Happy, Sad, Fear, and Neutral — is represented by distinct markers. The distribution reveals that happy responses (orange crosses) tend to have higher polarity values, indicating more positive sentiment, and exhibit a relatively high TTR, suggesting greater lexical diversity in positive speech. In contrast, angry and fearful responses (blue circles and red crosses) are more dispersed across the polarity spectrum, with many clustering in the negative range, reflecting their association with more negative sentiment. Neutral responses (purple diamonds) are spread across both low and high TTR values, indicating varied linguistic complexity in neutral expressions. This visualization highlights how emotional states influence both sentiment and lexical diversity, providing valuable insights into the linguistic patterns associated with different emotions.

\subsubsection{Result from Sentiment Polarity Across Different Emotion Responses}
\begin{figure}[ht]
    \centering
    \includegraphics[width=0.8\textwidth]{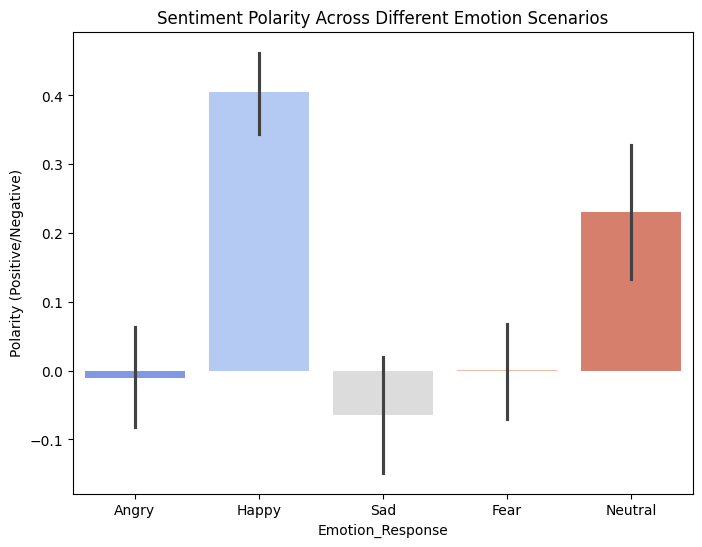} 
    \caption{Sentiment Polarity Across Different Emotion Scenarios. The bar plot illustrates the polarity (positive or negative sentiment) associated with each emotion category, including Angry, Happy, Sad, Fear, and Neutral responses. Happy responses show the highest positive polarity, while sad responses exhibit slightly negative polarity.}
    \label{fig:sentiment_polarity}
\end{figure}

The Figure~\ref{fig:sentiment_polarity} illustrates the Sentiment Polarity Across Different Emotion Responses, providing a clear understanding of how sentiment varies across emotions. The bar chart reveals that "Happy" emotions exhibit the highest positive polarity (~0.4), indicating a strong positive sentiment, whereas "Angry" emotions have the lowest polarity (close to -0.1), reflecting a more negative sentiment.
As expected, "Neutral" emotions remain close to zero, representing a balanced sentiment with neither strong positivity nor negativity. This figure effectively highlights the distinct variations in sentiment polarity across emotional responses, demonstrating that happiness is strongly associated with positive sentiment, while anger is linked to negative sentiment. These insights contribute to the development of emotion-aware VAs systems, enabling AI models to better recognize and respond to different emotional tones.
\section{Discussion}

Our findings provide key insights into how users respond to different emotional scenarios when interacting with emotion-aware voice assistants (VAs). By analyzing speech signals and linguistic content, we identified patterns in emotional responses, speech features, and sentiment polarity. These findings have several implications for the design of more effective and empathetic VAs, ensuring they provide more meaningful, natural, and human-like interactions.

\subsection{Emotional Response Patterns and User Strategies}

A key observation from our study is that users tend to adopt neutral emotional responses when interacting with negative emotional stimuli, such as anger, sadness, and fear. This suggests that individuals prefer a balanced, non-confrontational approach when faced with distressing scenarios, likely as a means of de-escalation. Such behavior aligns with established theories in interpersonal emotion regulation, where individuals employ strategies like cognitive reappraisal and expressive suppression to manage emotional interactions. Interestingly, while neutrality was the dominant response to negative emotions, participants demonstrated a greater emotional range when engaging with positive emotions, such as happiness. This variability suggests that users are more comfortable expressing emotions openly in positive contexts but may exhibit emotional restraint in negative situations. From a design perspective, VAs should recognize and mirror these behavioral tendencies, allowing for more intuitive interactions that do not feel forced or artificial. For VAs, this indicates that responses should be carefully calibrated to align with user tendencies. Instead of mirroring negative emotions or responding with excessive artificial empathy, VAs may be more effective when employing neutral yet supportive responses. Implementing context-aware, adaptive response strategies can help ensure appropriate engagement without overwhelming users.

\subsection{Speech Features as Indicators of Emotion}

Our analysis of speech signals, including root mean square (RMS), zero-crossing rate (ZCR), and speech jitter, highlighted significant variations between emotional states. Specifically, happy speech scenarios exhibited higher RMS and ZCR values compared to anger, fear, and sadness, indicating greater energy and pitch variation in positive emotional expressions. These findings underscore the potential of integrating advanced speech signal processing techniques into emotion-aware VAs. By continuously analyzing these speech features, VAs can enhance their real-time emotion recognition capabilities, allowing for more precise and adaptive responses to user emotions. Moreover, these insights could be leveraged to train VAs to detect the subtleties of human emotion, moving beyond basic categorical recognition (e.g., happy vs. sad) toward a more nuanced understanding of mixed and subtle emotional states.

\subsection{Linguistic Indicators of Emotion}

The word cloud and NLP analysis revealed distinct linguistic patterns associated with each emotional state. Angry responses featured terms related to conflict and frustration, while fearful responses included words indicating distress and a need for safety. Happy responses, in contrast, were characterized by words denoting enthusiasm and excitement. The correlation between sentiment polarity and type-token ratio (TTR) further illustrated that positive emotions tend to exhibit higher lexical diversity, while negative emotions are often expressed with a limited vocabulary. This suggests that individuals experiencing negative emotions may struggle to articulate their thoughts, potentially due to the cognitive load associated with distressing emotions. Conversely, positive emotions may encourage more elaborate and expressive language use. Understanding these linguistic variations can inform VA design, allowing for more tailored language models that adjust to user sentiment dynamically. Furthermore, beyond basic sentiment analysis, the integration of context-aware linguistic processing can enhance VA interactions. For example, VAs could analyze shifts in user tone and word choice over time, detecting emotional trends and adapting responses accordingly. This capability would make interactions feel more personalized and engaging, reinforcing user trust and reliance on these systems.

\subsection{Implications for Emotion-Aware Voice Assistant Design}

Our findings provide strong support for the integration of emotion-aware mechanisms in VAs. However, designing such systems requires careful consideration of cultural, gender, and contextual differences in emotional expression. Future implementations should prioritize:

\begin{itemize}
    \item \textbf{Adaptive Response Strategies}: VAs should dynamically adjust their level of empathy based on user emotional cues, ensuring responses are neither overly detached nor excessively emotional.
    \item \textbf{Multimodal Emotion Recognition}: Combining speech, linguistic, and contextual data can enhance emotion detection accuracy, mitigating issues related to accents, dialects, and background noise.
    \item \textbf{User-Centric Customization}: Allowing users to set personal preferences for VA responses can improve engagement and trust, ensuring that interactions align with individual comfort levels.
\end{itemize}

\subsection{Limitations and Future Work}

While our study provides valuable insights into emotional response strategies in human-VA interactions, several limitations should be acknowledged. First, the study was conducted in an online environment, which may not fully replicate real-world interactions with VAs. The controlled nature of the experiment could limit the generalizability of the findings to more dynamic, everyday scenarios. Additionally, cultural and linguistic diversity was not explicitly controlled, which may influence the applicability of our results across different populations. Emotional expression varies significantly across cultures, and future research should explore these variations to ensure that emotion-aware VAs are culturally sensitive and inclusive.
Another limitation is the reliance on single-dialogue interactions, which may not capture the complexity of users' emotional responses in prolonged or repeated engagements with VAs. Emotional dynamics often evolve over time, and future studies should investigate how users' emotional responses and strategies change during extended interactions. Furthermore, participants' first impressions of VA responses may have influenced their reactions, potentially limiting the scope of observed emotional dynamics. To address this, future research could incorporate iterative interactions, allowing participants to engage with VAs over multiple sessions.

To build on these findings, future work should focus on several key areas. First, cross-cultural studies are needed to better understand how emotional expression and regulation vary across different cultural contexts. This would help design VAs that are adaptable to diverse user needs and expectations. Second, long-term interaction studies could provide insights into how users' emotional responses and satisfaction evolve over time, particularly in scenarios where VAs are used regularly. Third, expanding the dataset to include more diverse user demographics and real-time conversational exchanges would improve the robustness and generalizability of emotion recognition models.
By addressing these challenges, future emotion-aware VAs can move closer to creating more natural, engaging, and emotionally intelligent human-AI interactions. The ultimate goal is not only to enhance the technical capabilities of VAs but also to ensure that they contribute positively to users' emotional well-being. Emotion-aware VAs should feel genuinely helpful, intuitive, and human-like, fostering trust and reliance in human-AI interactions.

\section{Conclusion}

This study explored human emotional response strategies in interactions with emotion-aware VAs using a role-swapping approach. Participants predominantly adopted neutral or positive emotional responses, particularly in negative scenarios, demonstrating a natural tendency toward emotional regulation. Speech feature analysis identified RMS, ZCR, and jitter as key indicators of emotional expression, while sentiment polarity and lexical diversity revealed distinct patterns across emotional states. These insights provide a foundation for developing adaptive, emotion-aware VAs that respond contextually and empathetically, improving user trust and engagement. Future research should focus on cross-cultural emotional models, deep learning-driven response adaptation, and multimodal emotion recognition to further enhance the emotional intelligence of VAs. By addressing these challenges, we can create VAs that not only enhance technical capabilities but also contribute positively to users' emotional well-being.

\begin{credits}
\subsubsection{\ackname} The authors extend their sincere gratitude to the anonymous reviewers for their valuable and constructive feedback. This research was funded by the Research Council of Norway under project number 326907. Additionally, support was provided by the Swedish Foundation for Strategic Research (SSF) through grant FUS21-0067.

\end{credits}
%
%
%
\bibliographystyle{splncs04}
\bibliography{mybibliography}

\begin{thebibliography}{10}
\providecommand{\url}[1]{\texttt{#1}}
\providecommand{\urlprefix}{URL }
\providecommand{\doi}[1]{https://doi.org/#1}

\bibitem{atta2024influence}
Atta, M.H.R., El-Gueneidy, M.M., Lachine, O.A.R.: The influence of an emotion
  regulation intervention on challenges in emotion regulation and cognitive
  strategies in patients with depression. BMC Psychology  \textbf{12}(1), ~496
  (2024)

\bibitem{barange2022impact}
Barange, M., Rasendrasoa, S., Bouabdelli, M., Saunier, J., Pauchet, A.: Impact
  of adaptive multimodal empathic behavior on the user interaction. In:
  Proceedings of the 22nd ACM International Conference on Intelligent Virtual
  Agents. pp.~1--8 (2022)

\bibitem{berking2012emotion}
Berking, M., Wupperman, P.: Emotion regulation and mental health: Recent
  findings, current challenges, and future directions. Current Opinion in
  Psychiatry  \textbf{25}(2),  128--134 (2012)

\bibitem{can2023approaches}
Can, Y.S., Mahesh, B., Andr{\'e}, E.: Approaches, applications, and challenges
  in physiological emotion recognition—a tutorial overview. Proceedings of
  the {IEEE}  (2023)

\bibitem{carolus2021alexa}
Carolus, A., Wienrich, C., T{\"o}rke, A., Friedel, T., Schwietering, C.,
  Sperzel, M.: ‘alexa, i feel for you!’ observers’ empathetic reactions
  towards a conversational agent. Frontiers in Computer Science  \textbf{3},
  682982 (2021)

\bibitem{el2011survey}
El~Ayadi, M., Kamel, M.S., Karray, F.: Survey on speech emotion recognition:
  Features, classification schemes, and databases. Pattern Recognition
  \textbf{44}(3),  572--587 (2011)

\bibitem{eyben2010opensmile}
Eyben, F., W{\"o}llmer, M., Schuller, B.: Open{SMILE}: The munich versatile and
  fast open-source audio feature extractor. In: Proceedings of the 18th ACM
  International Conference on Multimedia. pp. 1459--1462 (2010)

\bibitem{fischer2000relation}
Fischer, A.H., Manstead, A.S.R.: The relation between gender and emotions in
  different cultures. Gender and Emotion: Social Psychological Perspectives
  \textbf{1},  71--94 (2000)

\bibitem{fischer2004gender}
Fischer, A.H., Rodriguez~Mosquera, P.M., Van~Vianen, A.E.M., Manstead, A.S.R.:
  Gender and culture differences in emotion. Emotion  \textbf{4}(1), ~87 (2004)

\bibitem{gross2015emotion}
Gross, J.J.: Emotion regulation: Current status and future prospects.
  Psychological Inquiry  \textbf{26}(1),  1--26 (2015)

\bibitem{hoy2018alexa}
Hoy, M.B.: Alexa, siri, cortana, and more: An introduction to voice assistants.
  Medical Reference Services Quarterly  \textbf{37}(1),  81--88 (2018)

\bibitem{hu2022acoustically}
Hu, J., Huang, Y., Hu, X., Xu, Y.: The acoustically emotion-aware
  conversational agent with speech emotion recognition and empathetic
  responses. IEEE Transactions on Affective Computing  \textbf{14}(1),  17--30
  (2022)

\bibitem{huang2024relationship}
Huang, Z., Chen, S., Chen, H.: Relationship between emotional awareness and
  self-acceptance: The mediating role of emotion regulation strategies. Current
  Psychology pp.~1--9 (2024)

\bibitem{khalil2019speech}
Khalil, R.A., Jones, E., Babar, M.I., Jan, T., Zafar, M.H., Alhussain, T.:
  Speech emotion recognition using deep learning techniques: A review. IEEE
  Access  \textbf{7},  117327--117345 (2019)

\bibitem{kossack2023emotion}
Kossack, P., Unger, H.: Emotion-aware chatbots: Understanding, reacting, and
  adapting to human emotions in text conversations. In: International
  Conference on Autonomous Systems. pp. 158--175. Springer (2023)

\bibitem{kumar2024multimodal}
Kumar, C.U.O., Gowtham, N., Zakariah, M., Almazyad, A.: Multimodal emotion
  recognition using feature fusion: An {LLM}-based approach. IEEE Access
  (2024)

\bibitem{lieskovska2021review}
Lieskovsk{\'a}, E., Jakubec, M., Jarina, R., Chmul{\'\i}k, M.: A review on
  speech emotion recognition using deep learning and attention mechanism.
  Electronics  \textbf{10}(10), ~1163 (2021)

\bibitem{liu2022artificial}
Liu-Thompkins, Y., Okazaki, S., Li, H.: Artificial empathy in marketing
  interactions: Bridging the human-{AI} gap in affective and social customer
  experience. Journal of the Academy of Marketing Science  \textbf{50}(6),
  1198--1218 (2022)

\bibitem{ma2022emotion}
Ma, Y.: Emotion-Aware Voice Interfaces Based on Speech Signal Processing. Ph.D.
  thesis, LMU (2022)

\bibitem{ma2022should}
Ma, Y., Drewes, H., Butz, A.: How should voice assistants deal with users'
  emotions? arXiv preprint arXiv:2204.02212  (2022)

\bibitem{ma2024understanding}
Ma, Y., Nordberg, O.E., Zhang, Y., Rongve, A., Bachinski, M., Fjeld, M.:
  Understanding dementia speech: Towards an adaptive voice assistant for
  enhanced communication. In: Companion Proceedings of the 16th ACM SIGCHI
  Symposium on Engineering Interactive Computing Systems. pp. 15--21 (2024)

\bibitem{ma2023emotion}
Ma, Y., Zhang, Y., Bachinski, M., Fjeld, M.: Emotion-aware voice assistants:
  Design, implementation, and preliminary insights. In: Proceedings of the
  Eleventh International Symposium of Chinese {CHI}. pp. 527--532 (2023)

\bibitem{ma2024leveraging}
Ma, Z., Wu, W., Zheng, Z., Guo, Y., Chen, Q., Zhang, S., Chen, X.: Leveraging
  speech {PTM}, text {LLM}, and emotional {TTS} for speech emotion recognition.
  In: ICASSP 2024--2024 IEEE International Conference on Acoustics, Speech and
  Signal Processing ({ICASSP}). pp. 11146--11150. IEEE (2024)

\bibitem{mari2024empathic}
Mari, A., Mandelli, A., Algesheimer, R.: Empathic voice assistants: Enhancing
  consumer responses in voice commerce. Journal of Business Research
  \textbf{175},  114566 (2024)

\bibitem{mclean2019hey}
McLean, G., Osei-Frimpong, K.: Hey alexa... examine the variables influencing
  the use of artificial intelligent in-home voice assistants. Computers in
  Human Behavior  \textbf{99},  28--37 (2019)

\bibitem{mesquita1992cultural}
Mesquita, B., Frijda, N.H.: Cultural variations in emotions: A review.
  Psychological Bulletin  \textbf{112}(2), ~179 (1992)

\bibitem{ortony2022all}
Ortony, A.: Are all “basic emotions” emotions? a problem for the (basic)
  emotions construct. Perspectives on Psychological Science  \textbf{17}(1),
  41--61 (2022)

\bibitem{ortony1990s}
Ortony, A., Turner, T.J.: What's basic about basic emotions? Psychological
  Review  \textbf{97}(3), ~315 (1990)

\bibitem{Parvathi2025voice}
Parvathi, M., Pranathi, V., Varma, M., Satyanarayana, B.V.V.: Voice-based smart
  system for emotion recognition and regulation. In: Pareek, P., Mishra, S.,
  Reis, M.J.C.S., Gupta, N. (eds.) Cognitive Computing and Cyber Physical
  Systems. pp. 474--488. Springer Nature Switzerland, Cham (2025)

\bibitem{raamkumar2022empathetic}
Raamkumar, A.S., Yang, Y.: Empathetic conversational systems: A review of
  current advances, gaps, and opportunities. IEEE Transactions on Affective
  Computing  \textbf{14}(4),  2722--2739 (2022)

\bibitem{rathi2024analyzing}
Rathi, T., Tripathy, M.: Analyzing the influence of different speech data
  corpora and speech features on speech emotion recognition: A review. Speech
  Communication p. 103102 (2024)

\bibitem{srinivasan2022role}
Srinivasan, R., Gonz{\'a}lez, B.S.M.: The role of empathy for artificial
  intelligence accountability. Journal of Responsible Technology  \textbf{9},
  100021 (2022)

\bibitem{triantafyllopoulos2023overview}
Triantafyllopoulos, A., Schuller, B.W., {\.I}ymen, G., Sezgin, M., He, X.,
  Yang, Z., Tzirakis, P., Liu, S., Mertes, S., Andr{\'e}, E., et~al.: An
  overview of affective speech synthesis and conversion in the deep learning
  era. Proceedings of the {IEEE}  \textbf{111}(10),  1355--1381 (2023)

\bibitem{wani2021comprehensive}
Wani, T.M., Gunawan, T.S., Qadri, S.A.A., Kartiwi, M., Ambikairajah, E.: A
  comprehensive review of speech emotion recognition systems. IEEE Access
  \textbf{9},  47795--47814 (2021)

\bibitem{zaki2013interpersonal}
Zaki, J., Williams, W.C.: Interpersonal emotion regulation. Emotion
  \textbf{13}(5), ~803 (2013)

\end{thebibliography}
%




\end{document}